\documentclass[sigconf]{acmart} 

\usepackage{caption}
\usepackage{subcaption}

\usepackage{color,soul}
\setcopyright{none} 
\renewcommand\footnotetextcopyrightpermission[1]{} 
\begin{document}

\title{Evaluating Online Bandit Exploration In Large-Scale Recommender System} 

\author{Hongbo Guo, Ruben Naeff, Alex Nikulkov, Zheqing Zhu}
\affiliation{%
  \institution{Meta AI}
  \streetaddress{}
  \city{}
  \country{United States}}
\email{hongbog@meta.com, rubennaeff@meta.com, alexnik@meta.com, billzhu@meta.com}

\begin{abstract}
Bandit learning has been an increasingly popular design choice for recommender system. Despite the strong interest in bandit learning from the community, there remains multiple bottlenecks that prevent many bandit learning approaches from productionalization.  {One major bottleneck is how to test the effectiveness of bandit algorithm with fairness and without data leakage. Different from supervised learning algorithms, bandit learning algorithms emphasize greatly on the data collection process through their explorative nature. Such explorative behavior may induce unfair evaluation in a classic A/B test setting. In this work, we apply upper confidence bound (UCB) to our large scale short video recommender system and present a test framework for the production bandit learning life-cycle with a new set of metrics. Extensive experiment results show that our experiment design is able to fairly evaluate the performance of bandit learning in the recommender system.}

\end{abstract}


\maketitle
\pagestyle{plain}

\section{Introduction}
Short-form videos have got increasingly more traction on platforms such as on Instagram, Youtube and Tiktok, which demands an efficient recommender system (RS) that understands users' video interests with few interactions. 
A user's interactions to a recommended video are comprehensive, usually includes click, sharing, comments, likes or dislikes, video consumption time, etc. All of these objectives reflect the real user interest, and optimizing these objectives in a multi-task machine learning model (MTML) is one of the main-stream modeling choice for recommender systems. While supervised MTML methods have been widely applied to RS for exploitation, it remains a challenge how to explore for user's true preferences according to all signals from the multi-task setting \cite{covington2016deep, pi2020search, chen2019co, zhao2011integrating, gomez2015netflix, zhu2022scalable}. 
Multi-armed bandit (MAB) algorithms, as a means to explore users' true interests with minimum online interactions for recommender systems \cite{li2010contextual, chapelle2011empirical, zhou2020neural, xu2020neural, tang2015personalized, mcinerney2018explore, elena2021survey, silva2022multi, barraza2020introduction, mehrotra2020bandit}, are deemed as a suitable choice for such exploration problems. Despite the advancements in bandit learning for large scale recommender systems, critical issues still exist when applying bandit learning algorithms to the above setup, particularly difficulty of measurement in online tests.

 {In a traditional A/B test setting, a portion of users is allocated to control group, and another portion of users to test group where some new algorithm is applied on. This way, there is no data leakage between the two groups when we evaluate the new algorithm. 
However, this test setting does not work well for bandit algorithm in the recommendation system. The bandit algorithm's exploration nature may put disadvantage to the test group and induces unfair evaluation in this classic A/B test setting. 
More detailed, the test group has to sacrifice when conducting exploration. The control group does not explore nor sacrifice, however, it still gets benefit from the exploration of the test group. This is because the two groups share the same backbone MTML model trained on \emph{all users data}. This data contains the exploration information from the test group and allows control group to benefit unfairly.}


  {In this paper, we present an new A/B test setting for fair evaluation of  bandit algorithms in our production  recommender systems. In our test design, we train the MTML model separately for control and test groups, respectively using their own data. This test setting prevents the exploration information of test group from leaking to control group so that we get a fair evaluation.}
The rest of the paper is presented as follows. In section 2, we conduct literature review on bandit learning and its interactions with A/B testing. In section 3, we introduce the problem definition, and the bandit learning life-cycle. In section 4, we conduct a special A/B test for more accurate measurement, and introduce a framework of \textit{exploration efficiency} metrics that illustrate how our approach impacts the system's dynamics. Section 5 concludes the paper and briefly covers our future work.

\section{Related Work}
Multi-Armed Bandit (MAB) is a decision-making problem where the agent aims to learn the best action with minimum number of environment interactions where the actions do not present long-term consequences \cite{auer2002finite, audibert2009exploration}.  Various bandit learning algorithms have been developed to address the problem above, including UCB~\cite{auer2002finite, nguyen2019recommendation, lattimore2020bandit}, LinUCB \cite{li2010contextual, chu2011contextual}, neural network based algorithms \cite{zhou2020neural}, and epistemic neural recommendations algorithms \cite{zhu2023scalable}. There have been research that scales MAB to a Multi-Task learning setting that differs from ours. KMTL-UCB \cite{deshmukh2017multi} treats different contexts as different tasks and leverages context similarity to improve learning performance. AutoSeM \cite{guo2019autosem} leverages bandit learning for auxiliary task selection to improve model performance instead of directly exploring for target reward. \cite{soare2014multi} presents a theoretical analysis on multi-task linear bandits in a LinUCB setup. None of the multi-task settings above conducted online experiments on recommender systems. Beyond MAB, exploration and long-term value optimization are also critical in reinforcement learning based recommender systems \cite{xu2023optimize, zhu2023deep, chen2022off, chen2019top}.

\textbf{A/B Test Design for Bandit Learning Agents.} Few studies have been conducted on the fairness of A/B tests on bandit learning algorithms. It has been shown that bandit learning requires very careful user splitting in the A/B test, otherwise overlapping may violate the representativeness of the arms ~\cite{geng2020online}. A recent study combines A/B tests and MAB learning to achieve higher power in experiment design \cite{xiang2022multi}, which does not address the problem of exploration data leakage in the A/B testing setting.  {Beyond these studies, many of the causal inference communities' bandit learning works have been adaptive experimentation \cite{bakshy2018ae, burtini2015survey, rafferty2019statistical, geng2021comparison}, which focuses on using bandit learning to choose experiment treatment groups and they are different from our setting. }

\section{Problem, Solution and System Set Up}

\subsection{Formulation of the problem}
We frame recommender system design as a non-parameterized multi-armed bandit (MAB) problem where an agent gets an observation at time step $t$, takes an action and receives a reward at time $t+1$. More formally, we define the recommender MAB problem as the following:

\textbf{Action Space $\mathcal{A}$}:  
   {In our application, videos are clustered by video features and each cluster corresponds to one topic. Here each topic is a bandit arm. This clustering helps to reduce the action space \cite{shams2021cluster, gentile2017context, pandey2007multi}. Though our non-parametric bandit solution is not a contextual bandit model with trainable weights\footnote{(Such as LinUCB~\cite{li2010contextual} which takes features as input and predicts UCB scores).}, some contextual features of the videos are indeed utilized in our system. Conceptually, videos belonging to the same cluster are samples of the same bandit arm, and they share their features thanks to the clustering. } The action is to choose one of the topics. 
The size of action space is $\left| \mathcal{A} \right| = K$, representing the $K$ different topics and $K$ is limited to a few hundreds.


\textbf{Observation Space $\mathcal{S}$}: There are two attributes in the observation space for recommender system, user and available actions. Although we do not leverage parameterization for modeling uncertainty, we adopt different uncertainty calculation for different users. Each <user, topic> pair has its own UCB score calculated. Also in production recommender systems, not all recommendations in the action space is available at time step $t$. More formally, we define $S_t = (C_t, \mathcal{A}_t)$, where $C_t$ is the user and $\mathcal{A}_t \in \mathcal{A}$ is the subset of available actions. Given the definitions above, at round $t+1$, the agent selects a best action $a_{t+1}$ from the action space $a_{t+1} \in \mathcal{A}_t$. 

\textbf{Reward Space}: The reward $r_a(T, C_{t-1})$ is the feedback given by the environment once action $a$ is taken for task $T$ and user $C_{t-1}$. Similar to many other recommendation systems such as Netflix or Spotify ~\cite{gomez2015netflix, mcinerney2018explore, mehrotra2020bandit}, the tasks of our recommender system include users play, comment, share or like.

\subsection{UCB applied to MTML Recommender System}
In this work, we adopt the optimism facing uncertainty principle and apply Upper Confidence Bound (UCB) approach for bandit learning and execution to the recommender system. 
The UCB formula adopted in our experiment consists of two items, the expected reward of an action applied to a context and the exploration bonus: 
\begin{equation}
  U_{t, a} (C_{t-1}) = \sum_{T\in \mathcal{T}}\alpha_T\hat{r}_a(T, C_{t-1}) + \gamma \sqrt{\frac{\ln N_{t}(C_{t-1})}{N_{t,a} (C_{t-1})}}.
  \label{eq:ucb}
\end{equation}
Here $a$ represents the arm or topic and $\mathcal{T}$ is the set of tasks. $N_t(C_{t-1})$ represents the total number of times the recommender system interacts with the user $C_{t-1}$ and $N_{t, a}(C_{t-1})$ represents the number of times the recommender system has recommended topic $a$ to the user $C_{t-1}$. $\gamma$ is a hyperparameter to adjust the aggressiveness for exploration and $\alpha_T$ is the importance of task $T$.

The goal of this UCB formula is to induce a more explorative behavior when uncertainty is high.
Assuming the task of interest here is to predict whether user will click the video. A click is a good reward, and a non-click is a bad reward. { For example, if we tried 100 times on football arm and got 99 clicks, then we want to keep recommending football videos so as to exploit. Besides football, the user may also be interested in another topic, such as the romantic videos. On romantic arm, lets say we had recommended 3 videos and user clicked 1 of them.  It looks that romantic is not favorable, but we only tried as few as 3 times after all. It may be worth taking some risk and explore more on romantic since we are uncertain about it due to limited number of trials.
}

\begin{figure}
  \centering
  \includegraphics[width=0.6\linewidth]{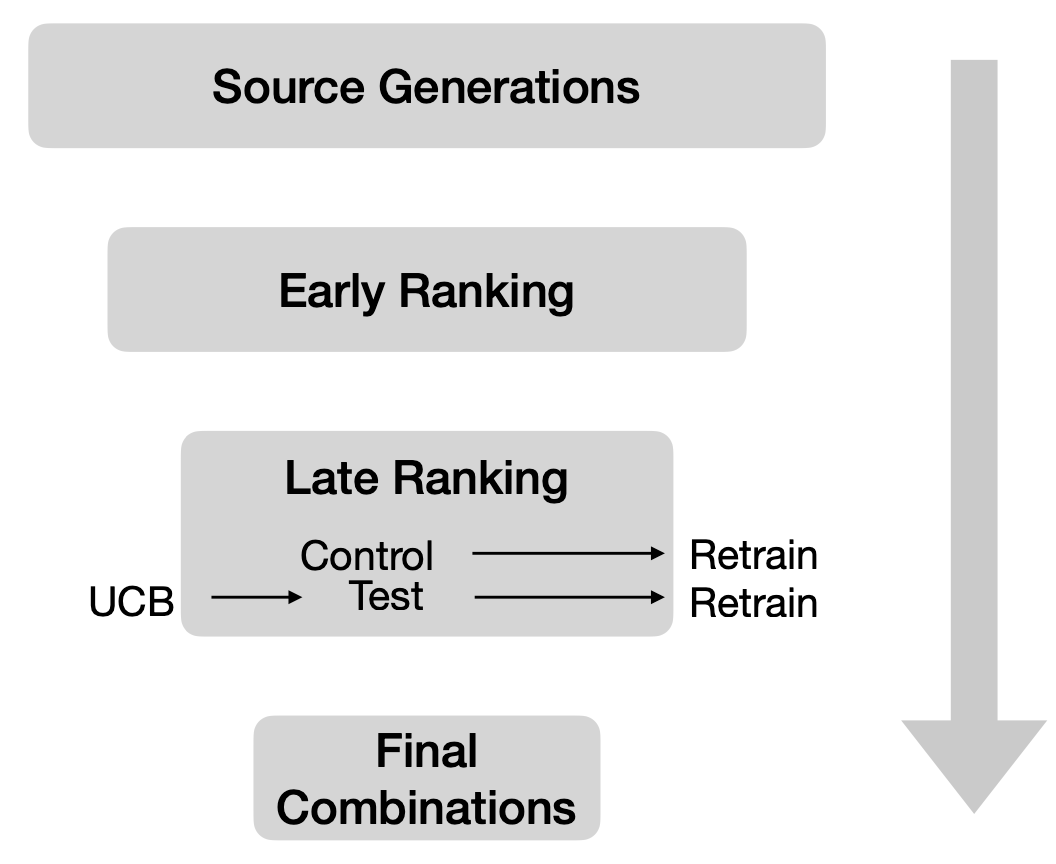}
  \caption{Bandit Learning for Ranking Funnel}
  \label{fig:RS}
\end{figure}

 {The position of UCB exploration in our recommender system is depicted in Fig.~\ref{fig:RS}. 
Our recommender system is a complicated multi-stage funnel system, which consists of content source generation, early ranking and late ranking for multiple tasks. This MTML is a very powerful supervised model and the late ranking score serves as a very good estimation of reward (the 1st item of of Eq. \ref{eq:ucb}). We add the UCB exploration bonus (the 2nd item of Eq. \ref{eq:ucb}) to the late ranking score. This contextual non-parametric version of UCB offers an elegant estimation of the confidence set for all the tasks and hence simultaneously explores and learns to personalize for multiple different tasks such as click, comment, like and share. 
}

\begin{figure}
  \centering
  \includegraphics[width=1.1\linewidth]{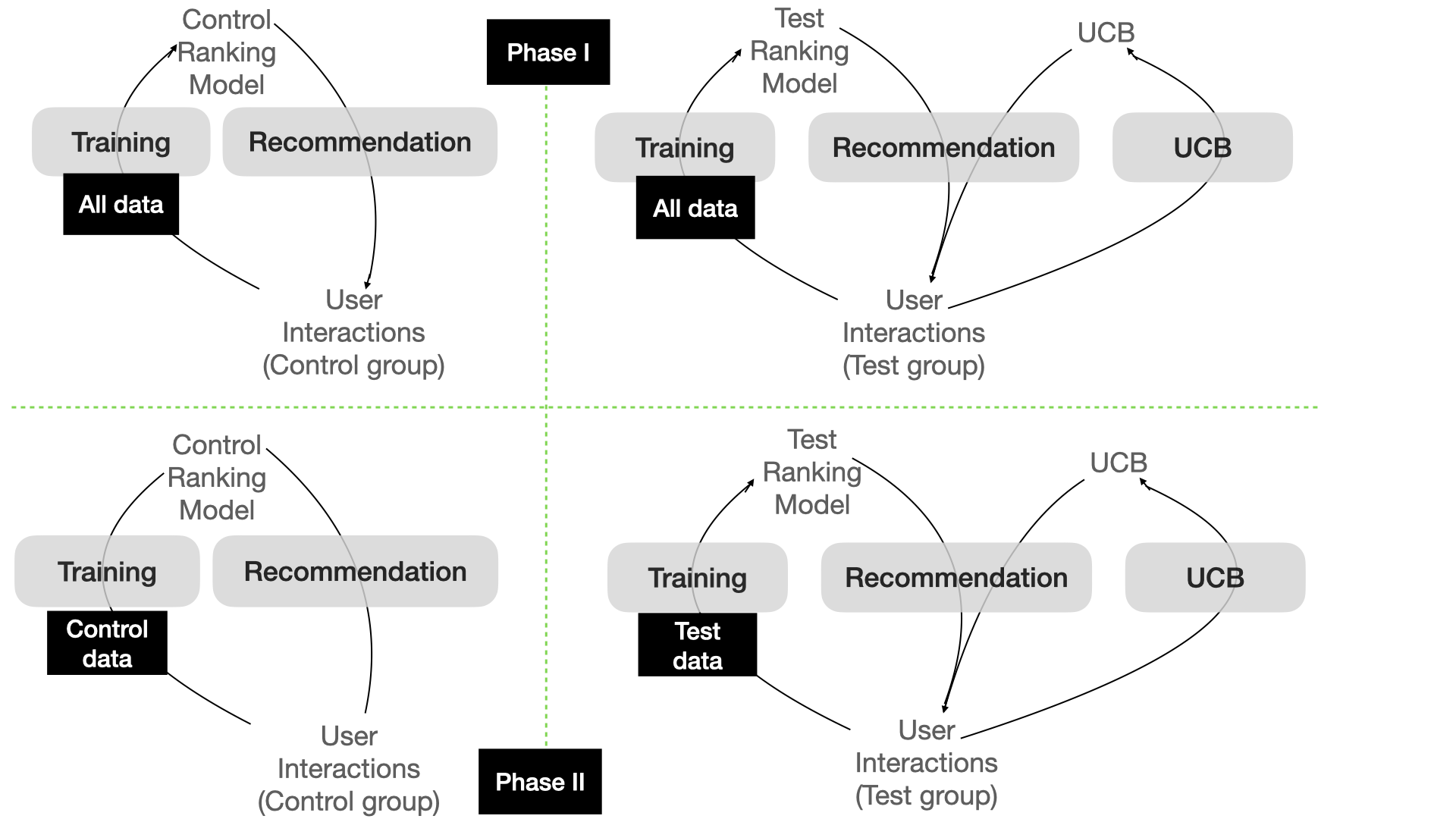}
  \caption{Bandit Learning A/B Test Phases}
  \label{fig:control_test}
\end{figure}

\subsection{Measure Bandit Learning in A/B Test}

Since UCB is an online bandit learning algorithm, it introduces natural complexity when trying to apply hypothesis testing methods, such as A/B testing, to measure the effectiveness.  {The bandit learning agent in the test group sacrifices its immediate performance to explore and learn new information from the user. However, this valuable information is actually leaked to the control group's agent due to the fact that all groups share the same backbone MTML. This yields unfairness to the test group's agent because 
both groups' agents benefit from this valuable information but the bandit learning agent in the test group has to sacrifice its performance to retrieve such information.} 

To address the issue above, we adopted a customized A/B testing mechanism to bound the performance of bandit learning algorithms. Control and test groups are of same size and each consists of a portion of all users. Note control and test group each is huge enough to allow the models to generalize well. Instead of a single phase A/B test with all data shared for model refreshing, we will set up the experiments for two phases. 

In Phase I, we test the effect of UCB without dedicated posterior update in each group to measure the lower bound of bandit learning capability. In this phase, the uncertainty of UCB calculated in 2nd item of Eq. \ref{eq:ucb} is applied to the test group, but not to the control group. All groups share the same reward mean estimation $\hat{r}_a(T, C)$ which is from MTML model trained on all data across all groups. Given that control and test groups share the same MTML, we expect the test group to personalize better first,  but will be caught up with by the control group over the long term due to the data leakage.

 {In Phase II, we enable data split between test and control group and train dedicated MTML models, respectively. The control group's MTML is trained by its own users data, and so is the test group. Their training data are of the same amount  since control and test groups are of same population size. } This allows a proper posterior update for each group separately, without data leakage. We expect the performance of the test group to improve on top of phase I's result with better exploration strategy. Phase I and Phase II together provides the upper and lower bounds of the bandit learning performance.  {For more details, please refer to Figure \ref{fig:control_test} where upper half of the figure is Phase I and lower half is Phase II, while left half is control and right half is test.}

\section{Analysis}
\subsection{Set Up}
We allocate some users into our experiment and divide them into two groups: the control group and test group. They are of the same population size, and each group size is huge due to the fact that on our platform there are over one billion users. Our experiment ran in two phases lasting for 3 and 2 weeks, respectively. In Phase I, we applied UCB to the test group, retraining our model with the full data set collected across all users. All groups are exposed to the same MTML ranking model. In Phase II, we retrained the ranking model on each group separately. Each of the control and test groups contains large amount of data. Thus, each group is large enough to allow the ranking model to be well trained. In both phases, we trained the model daily. 

\subsection{Metrics}
To measure the success of our approach, we look at the number of \emph{plays} of short-form videos. To measure the quality of those plays, we also look their \emph{loop rate} and \emph{skip rate}, i.e., if users watch a video repetitively, 
for less than a few seconds. 

In addition, to evaluate whether our policy follows a near-optimal posterior belief, we develop a suite of \emph{exploration efficiency} metrics. Using a coarse taxonomy of ~30 topics, we create posterior distributions for \(p(\text{relevant})\) for each <user, topic> pair, based on past recommendations, where we mark the completed videos as relevant. Note that this includes never-seen topics.\footnote{We use beta distributions with a light prior based on topic's and user's averages.} By sampling from these posteriors, we can compute the probability that a topic would be the most relevant, i.e., we create a probability distribution $P$ over all topics. Ideally, our policy follows this belief, so we compare this with our recommended topic distribution $Q$.\footnote{We compute $P$ with data before today, while $Q$ is computed with today's data.} For these metrics, we assume independence between topics.

Now we can define the \emph{exploration inefficiency (EI)} as the KL divergence between $P$ and $Q$, \emph{topic diversity (TD)} as the number of recommended topics, 
\emph{interest uncertainty (IU)} as the entropy of $P$, and \emph{topic excellence (TE)} as the probability that any given recommendation happens to be on the most relevant topic.
\begin{align}
    EI &= D_{\text{KL}}(P\parallel Q) = \sum_{i} p_i \log \frac{p_i}{q_i} \\
    TD &= |Q| = \text{\# topics} \\
    IU &= H(P) = - \sum_{i} p_i \log p_i  \\
    TE &= P \cdot Q = \sum_i p_i q_i
\end{align}

We average these user-level metrics across all users. For topic excellence, the average is weighted by the number of plays. In the KL divergence, we impute missing topics with a small value, which skews the metric worse for light users. Our experiment aims for a lower exploration inefficiency and a higher topic excellence.

\begin{figure}
  \centering
  \includegraphics[width=.9 \linewidth]{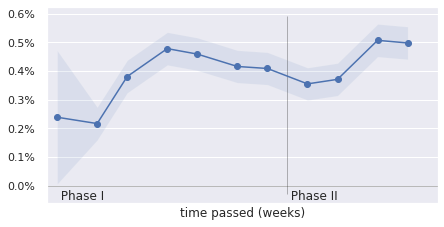} 
  \caption{Effect on play count (test/control-1)}
  \label{fig:intentional_plays}
\end{figure}

Comparing treatment with control, we find an increase bounded around +0.5\% plays of short-form videos (Fig. ~\ref{fig:intentional_plays}). We make three observations: plays go up in phase I, yet after a while these gains start to diminish; then in phase II we see again an increase. We will discuss some interpretations later on.

\begin{figure}
     \centering
     \begin{subfigure}[b]{0.48\linewidth}
         \centering
         \includegraphics[width=\textwidth]{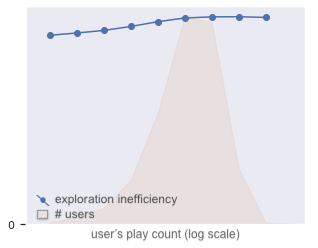}
         \caption{exploration inefficiency}
         \label{fig:phase0_exploration_inefficiency}
     \end{subfigure}
     \begin{subfigure}[b]{0.48\linewidth}
         \centering
          \includegraphics[width=\textwidth]{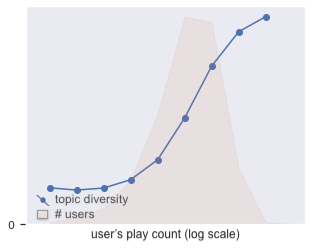}
         \caption{diversity (\# topics)}
         \label{fig:phase0_num_topics_seen}
     \end{subfigure}
     \begin{subfigure}[b]{0.48\linewidth}
         \centering
          \includegraphics[width=\textwidth]{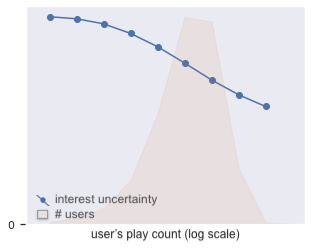}
            \caption{interest uncertainty}
            \label{fig:phase0_interest_uncertainty}
         \end{subfigure}
     \begin{subfigure}[b]{0.48\linewidth}
         \centering
          \includegraphics[width=\textwidth]{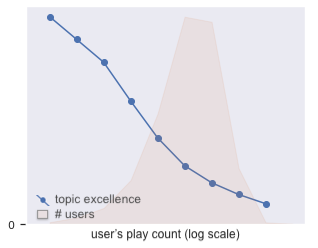}
                \caption{topic excellence}
                \label{fig:topic_excellence}
         \end{subfigure}
\caption{Exploration efficiency by user activity (control)}
\label{fig:phase0_exploration_efficiency}
\end{figure}

In our control group (Fig. ~\ref{fig:phase0_exploration_efficiency}), we see the effects of a classic \emph{feedback loop problem}: for more active users we are exploring less efficiently, and while we get a clearer picture of which topics have potential, we do not seem to be exploring them. The shaded bell curve is the distribution of total plays per user.


\begin{figure}
     \centering
     \begin{subfigure}[b]{0.48\linewidth}
         \centering
         \includegraphics[width=\textwidth]{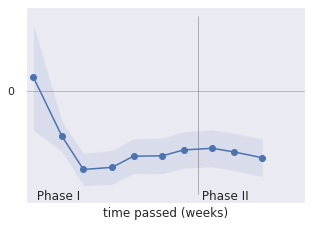}
         \caption{exploration inefficiency}
         \label{fig:exploration_inefficiency}
     \end{subfigure}
     \begin{subfigure}[b]{0.48\linewidth}
         \centering
          \includegraphics[width=\textwidth]{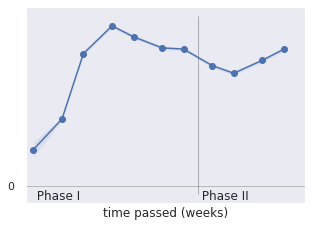}
         \caption{diversity (\# topics)}
         \label{fig:num_topics_seen}
     \end{subfigure}
     \begin{subfigure}[b]{0.48\linewidth}
         \centering
          \includegraphics[width=\textwidth]{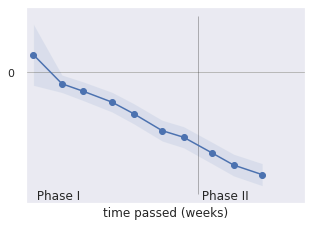}
            \caption{interest uncertainty}
            \label{fig:interest_uncertainty}
         \end{subfigure}
     \begin{subfigure}[b]{0.48\linewidth}
         \centering
          \includegraphics[width=\textwidth]{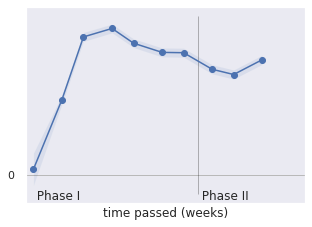}
                \caption{topic excellence}
                \label{fig:topic_excellence}
         \end{subfigure}
        \caption{Effect on exploration efficiency (test/control-1)}
        \label{fig:exploration_efficiency}
\end{figure}

In Fig. ~\ref{fig:exploration_efficiency}, we see that our experiment indeed decreases our exploration inefficiency, which stays consistently low throughout Phase I and II. As a result, diversity increases, as uncertain topics are promoted. Yet, as certainty grows, these promotions get weaker. In Phase II, when our model learns, diversity grows again, as new interests are recommended. 
Topic excellence also shows
an initial jump as the potential of our unexplored topics is high (we're optimistic), steadily decreasing as we face reality, and, in Phase II, increasing when we recommend our newly discovered and superior topics again. To recap, exploration inefficiency dropped, and topic excellence grew, as we had intended.

\begin{figure}
    \centering
    \begin{subfigure}[b]{0.48\linewidth}
        \centering
        \includegraphics[width=\textwidth]{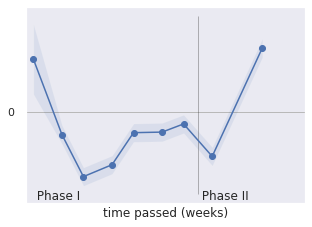}
        \caption{loop rate}
        \label{fig:exploration_inefficiency}
    \end{subfigure}
    \begin{subfigure}[b]{0.48\linewidth}
        \centering
        \includegraphics[width=\textwidth]{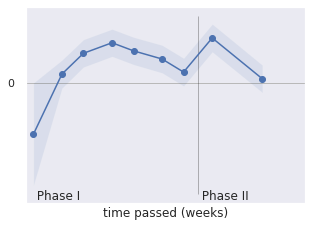}
        \caption{skip rate}
        \label{fig:num_topics_seen}
    \end{subfigure}
    \caption{Effect on video engagement (test/control-1)}
    \label{fig:engagement}     
\end{figure}

In Fig. ~\ref{fig:engagement}, 
the initial bump might point at novelty effects, as new topics are boosted. Then rates reverse, suggesting that users have the opportunity to explore new interests by swiping through some of the suggestions. In Phase II, when the model learns from the newly-explored topics, users seem to have found their interests: loop rates are higher than in control and skip rates end up statistically tied to control.

\section{Conclusion and future work} 
In this work, we formulate our short-form video recommender system as a contextual non-parametric multi-armed bandit problem. We introduce a practical bandit learning lifecycle for recommender systems, and present a customized A/B test design for fair measurement of performance.   {The test is a multi-stages A/B testing framework, and it provides more experimental information than classical A/B testing. It is able to fairly evaluates bandit learning performance without data leakage.  At last, our experiments show empirically that the bandit learning approach significantly improves personalization for multiple tasks. These observations had been testified by our customized A/B testing framework and comprehensive metrics.}

\bibliographystyle{ACM-Reference-Format}
\bibliography{ucb-paper}

\end{document}